# Population Balances in Case of Crossing Characteristic Curves: Application to T-Cells Immune Response

Qasim Ali[1,3], Ali Elkamel[2], Frédéric Gruy[3], Claude Lambert[4], and Eric Touboul[5]

[1] Department of Mathematical Biology, University of Western Ontario, London, ON, Canada
[2] Department of Chemical Engineering, University of Waterloo, Waterloo, ON, Canada
[3] Center for Chemical Engineering, ENSM – Saint Etienne, 158 cours fauriel, ENSM-SE, Saint Etienne, France
[4] CHU – Saint Etienne, Immunology Lab., University Hospital, F 42055, Saint-Etienne, France
[5] Henry Fayol Institute, ENSM – Saint Etienne, 158 cours fauriel, ENSM-SE – Saint Etienne, France

## Abstract

The progression of a cell population where each individual is characterized by the value of an internal variable varying with time (e.g. size, mass, and protein concentration) is typically modelled by a population balance equation, a first-order linear hyperbolic partial differential equation. The characteristics described by internal variables usually vary monotonically with the passage of time. A particular difficulty appears when the characteristic curves exhibit different slopes from each other and therefore cross each other at certain times. In particular, such a crossing phenomenon occurs during T-cell immune response when the concentrations of protein expressions depend upon each other and also when some global protein (e.g. Interleukin signals) that is shared by all T-cells is involved. At these crossing points, the linear advection equation is not possible by using the hyperbolic conservation laws in a classical way. Therefore, a new transport method is introduced in this article that is able to find the population density function for such processes. A multi-scale mathematical modelling framework is employed. At the first scale, two processes, which are termed as the activation and reaction terms (assimilated as nucleation and growth in chemical processes), are studied as independent processes. At the second scale, the dynamic variation in the population density of activated T-cells is investigated in the presence of activation and reaction terms. The newly-developed transport method is shown to work in the case of crossing and to provide a smooth solution at the crossing points in contrast to the classical PDF techniques.

INTRODUCTION

Deterministic models, like continuum and ensemble models, work either at the single cell level or at the population level. We need a modelling technique that can work at both scales to provide a relation between both levels.[1,2] Population balance models (PBMs) have provided such an interface which connects the different scales and deals with several particulate processes simultaneously.[3,4]

Population balance modelling is a very useful tool in chemical engineering.[5] It has diverse applications including polymerization, crystallization, precipitation, and several other chemical industrial processes.[6,7] In the biological sciences, PBM has been employed heavily due to vast applications of population dynamics in the modelling of continuous variation in the microbial populations,[2,8,9] phenomena such as cell growth,[10] intracellular reactions and division with stochastic partitioning,[11–13] and gene regulatory processes by population mediated signalling.[14]

Mathematical modelling for the concentrations of multi proteins in a given T-cell has been investigated by several researchers.[15–18] These models were cast as ordinary differential equations by representing the kinetics of the biochemical reactions.[19–22] A more sophisticated approach considers the dynamics of cell population, where each cell is characterized by the concentration of one or several proteins. This leads to a multi-dimensional population balance equation with the order depending upon the number of internal variables (protein concentrations).[2,23]

Each term in the population balance equation (PBE) describes a particulate process.[4,5] A general PBE is an integro-partial differential equation (integro-PDE) depending upon the spatial and temporal scales which cannot be solved analytically. Numerical methods for solving PBE[24–26] or stochastic approaches[12,27,28] have to be implemented to get the population density function.

Particular to our problem, the PBE with growth and nucleation processes is represented by a linear hyperbolic PDE with a source term. The temporal change in the internal parameters (variables) draws the characteristic curves along which the whole population travels.[8,29] In the literature, the flux of the population was unique. Therefore, these types of equations were solved analytically and numerically with high accuracy because the growth rates were considered either constant or linear.[30] In other words, the characteristics were showing monotonic behaviour with a translational phenomenon.[31,32]

However, in the case of Ostwald ripening,[33] particles with intermediate sizes begin to grow and then dissolve, leading to non-monotonous characteristic curves.[34] In another work, PBM was discussed with a non-monotonous growth process (termed a reaction process) and nucleation process (termed an activation process) in biochemical engineering[35] when the internal variable (concentration of protein) showed non-monotonous behaviour versus time. However, the characteristic curves followed only the translational phenomenon (i.e. the slopes of all the concentration curves were the same). Therefore, it was possible to write coupled PDEs to model the problem.

Such dynamical systems are frequently studied experimentally using the Flow-Cytometry technique which provides the population density of cells with respect to their associated one or more surface proteins concentrations like CD3, CD25, and CD69.[15,21,36] In particular the crossing phenomenon occurs during T-cells immune response when the concentrations of protein expressions depend upon each other and some global protein (e.g. Interleukin signals) is also involved which is shared by all T-cells.[37,38]

The present article describes an application when such characteristic curves are non-monotonic and each curve follows its own path (i.e. the slope is different for each curve) and thus the curves cross each other. Therefore, the flux at the crossing points does not remain unique. Since each curve has different slope, it is not possible to deal with such a problem using the classical conservation laws (such as PBE).

LINEAR HYPERBOLIC CONSERVATION LAWS

**Classical PBE with Growth and Nucleation**

Let us define $x$ as any internal variable representing the characteristic (size, mass, concentration, etc.) of individuals that vary with the passage of time. We suppose that the characteristic of the individual begins to vary at time $t = 0$ and does so continuously during the whole simulation time, i.e. the actual time $t$. In a similar way, we assume that at each time step $t$ there is some new individual beginning to vary its characteristics which will continue the variation during the rest of the actual time $t$. This means that two independent time variables are involved: the time at which an individual starts varying its characteristic, the 'starting time' $\tau$; and the actual time $t$ in which the individual continues the characteristic variation. Thus, if $x$ is the characteristic of individuals then we get a family of curves $x(t, \tau)$, the characteristic curves, as shown in Figure 1. For each individual, and for a fixed $\tau$, $x(t,\tau)$ is classically derived from a kinetic law:

$$\begin{cases} \dfrac{dx(t,\tau)}{dt} = g(x,t) & \text{for a given } \tau \text{ and } t > \tau \\ \text{with initial condition} \quad x(t,\tau) = x_0 & \text{for } t \leq \tau \end{cases} \quad (1)$$

This represents $g(x,t)$ as the growth rate defined as the slope of internal variable $x$. Let us define the nucleation term as the source term $N(x,t)$, which varies between $x$ to $x+\Delta x$ within time $t$ to $t+\Delta t$. For this, consider that $B(t)$ is the nucleation rate at a fixed starting time $t = \tau$ and $D(x)$ is the initial distribution of the population over internal variable $x$. For simplicity, we consider that $D(x) = \delta(x - x_0)$. It is then possible to write the source term as follows:

$$N(x,t) = B(t)\delta(x - x_0).$$

Following the linear hyperbolic conservation laws, we can write the population balance equation as follows:[39]

$$\begin{cases} \dfrac{\partial n(x,t)}{\partial t} + \dfrac{\partial \big(g(x,t)n(x,t)\big)}{\partial x} = 0 \\ n(x_0,t) = n_0(t) \end{cases} \quad (2)$$

where the initial condition is $n(x,0) = n(x)$ and the boundary condition is $n_0(t) = B(t)/g(x_0,t)$. This is a simple advection equation with a source term that is represented as a boundary condition. The number of individuals (population) activated at time $\tau$ are transported along the characteristics and remain the same throughout the simulation time $t$; however, the population density $n(x,t)$ is not constant along the characteristics. The variation of $n(x,t)$ along these curves can be found by using the total derivative of Equation (2):

$$\dfrac{dn(x(t),t)}{dt} = -n(x(t),t)\dfrac{\partial g}{\partial x}(x(t),t) \quad (3)$$

The characteristic curves provide a useful way to analyze Equation (2). In the framework of hyperbolic equations, these problems are countlessly studied as advection equations when the growth rate $g(x,t)$ is either a constant or linear with $x$. In such cases there was no crossing involved, and therefore the flux on each characteristic curve was unique. However, for crossing curves when each characteristic has a different slope, the advection equation cannot be modelled at the crossing points because the flux is no longer unique. Of course, there is no use of classical numerical/analytical methods when Equation (2) does not represent the physical phenomenon at crossing points. This phenomenon is briefly explained in the following section.

**Crossing Characteristics**

Let us consider two populations, $P_A$ and $P_B$, and follow their respective rates of change. Population $P_A$ starts the dynamical process at time $t = \tau_A$ and varies its characteristics less rapidly than the variation in characteristics $x_B = x(\tau_B, t)$ of the population $P_B$ that starts the process at $t = \tau_B$ as shown in Figure 2. Due to such variation in the characteristics of $P_A$ and $P_B$, the curves can cross each other between $x$ and $x + \Delta x$ (when $x_A = x_B$) at a particular time between $t$ and $t + \Delta t$. Certainly, the populations travelling along these characteristic curves will increase the total population at this point. By finding the population densities for the two populations, we can simply add them to find the total population density at this time $t$, i.e. $n(x,t) = n(x_A, t) + n(x_B, t)$.

**Transport Method Formulation**

The transport method (TM) obeys the conservation laws and allows us to find the analytical solution by following the behaviour of the characteristic curves. The method is independent of the complexity of the characteristic curves as it follows the initial population of the activated T-cells to find the population density at any required time $t$.

Let us discretize the $\tau$-axis into a grid $\tau_i$ with a step-size $\Delta \tau$, and the $t$–axis into a grid $t_j$, with a step-size $\Delta t$. Consider the initial population during the time $t \in [\tau_i, \tau_i + \Delta \tau]$ as demonstrated in Figure 3. The total number of individuals starting their dynamical process at $t \in [\tau_i, \tau_i + \Delta \tau]$ can be approximated as follows:

$$\int_{\tau_i}^{\tau_i + \Delta \tau} B(\tau) d\tau \approx B(\tau_{i+\frac{1}{2}}) \Delta \tau \quad \forall i = 1, 2, ... \tag{4}$$

We find the population density for $t > \tau_i$ by following the curves. If we observe the flow of population on the internal variable axis ($x$ – axis), it seems like the population is distributing over $x$. By linear hyperbolic conservation laws, it can be deduced that the population $B(\tau)$, began its dynamical process at starting time $\tau_i$ to $\tau_i + \Delta\tau$, varies its characteristic (internal variable) simultaneously in each step size at the rate equal to its reaction (growth) rate $g(x,t)$ as defined in Equation (1). Therefore, the population remains the same during the whole simulation time $t$. If $n(x,t)$ is the population density, then the total population within the interval $[x_i, x_i + \Delta x_i]$ at any time $t$ would be equal to Equation (4):

$$n(x_i, t) \cdot \Delta x_i = B(\tau_{i+\frac{1}{2}}) \Delta\tau \quad \forall i = 1, 2, \ldots \tag{5}$$

Note that we use the same index '$i$' for internal variable $x$ and for starting time $\tau$. This relation expresses that the density of the individuals in each step size $\Delta x_i$ is same as the number of individuals in step size $\Delta\tau$. In addition, the ratio between the variable step size $\Delta x_i$ of the curve $x$ and the time step $\Delta\tau$ ($=\Delta t$) provides the slope of $x(t,\tau)$ that should be equal to $g(x,t)$ defined in Equation (1). If the slopes of all characteristic curves are same such that the curve $x(t,\tau_A) = x(t,\tau_B)$ for each $\tau_A \neq \tau_B$ then we can write the curves as $x(t,\tau) = x(t-\tau)$ which shows that each curve is the translation of the curve starting at $\tau = 0$. This allows us to assimilate the slope of the curves $x(t,\tau)$, i.e. $g(x,t)$, as the growth rate of the PBE, i.e. hyperbolic PDE. These types of characteristic curves are well-studied in the literature of hyperbolic conservation laws and can be very accurately approximated using any numerical scheme, particularly when these curves are monotonic. For non-monotonic curves when each curve is the translation of the first curve, i.e. at $\tau = 0$, there is a possibility of finding the population density function by using a conservation equation, splitting the characteristics at the local extrema, and finding the population density by making separate PDE for each split.[19,35] Note that the splitting is done to avoid the crossing in the characteristic curves.

There is another possibility when each characteristic curve has a unique slope such that the curve $x(t,\tau_A) \neq x(t,\tau_B)$ for each $\tau_A \neq \tau_B$, as discussed in the subsection above, then the curves $x(t,\tau) \neq x(t-\tau)$. Therefore, the slope of each curve $x(t,\tau)$ needs to be represented by a unique function $g_\tau(x,t)$, for each $\tau$. In this case, it is still possible to find the PDE and to solve it by numerical methods if the curves do not cross each other. There could be some approximation error because we must fix the grid (possibly by interpolation) for all characteristic curves despite each characteristic having a unique set of points (and therefore each should have a unique grid). However, for crossing characteristics, it is not possible to write the conservation equation since the population at the crossing region has more than one characteristic at each time $t$ while the PDE requires a single characteristic passing through each point. Splitting the curves is also not possible since each curve can have its own local extrema that may be different from others. This phenomenon is considered in the sections below.

The idea of TM is close to the MOC, when $x(t, \tau) = x(t - \tau)$, since a set of individuals is followed along a "characteristic surface" $x(t,\tau)$ (i.e. in 3 dimensions), and then a projection on the $x$-axis is performed (i.e. in 2 dimensions). However, the classical characteristic method MOC cannot be followed in the case of $x(t, \tau) \neq x(t - \tau)$ since it is not possible to write the conservation equation (Equation (2)), contrary to the transport formulation which allows us to find the population density by using the basic idea behind conservation law (without using the conservation equation) and writes it in a simple form instead of using PDEs. Following are the two examples quoted to illustrate the method in the case of crossing characteristics.

*Example 1: Monotonous case*

This example shows the presence of discontinuities in the density function $n$ even when the slopes of the characteristics vary continuously. For example, if the value $x_0$ is attained by a single characteristic up to time $t_0$, and then by two characteristics, the density of population at $x_0$ will show a jump at time $t_0$, because at this time an additional population is considered. We define the growth function as follows:

$$g_\tau\left(x(t,\tau),t\right) = \begin{cases} 1 & \text{if } \tau < 3 \\ 1+(\tau-3) & \text{if } \tau \geq 3 \end{cases}$$

Note that the slopes of the characteristic curves begin to continuously increase from $\tau = 3$. For example, at time $t = 6$, there are 3 zones along the $x$-axis (see Figure 4a). The first zone is $0 < x < 3$, the second zone is $3 < x < 4$, and the third zone is $4 < x < 6$. The starting rate, $B(t)$, is constant and equal to 1. Although we do not need the initial and boundary conditions since TM does not use the neighbouring values, we can define them according to given conditions of the growth rate $g_\tau(x_0,t)$ and starting rate $B(t)$. Taking the ratio between the starting rate $B(t)$ and the growth rate at the initial level $g_\tau(x(t,\tau),t)$ provides us the boundary condition $n(x_0,t) = B/g_\tau(x_0,t)$ with initial condition $n(x,0) = 0$ since there is no initial distribution defined over the internal variable $x$. The actual time step size $\Delta t = 0.02$ is the same as the starting time step size $\Delta \tau$ while the internal variable $x$-step size varies as $\Delta x(t_i,\tau_j) = x(t_{i+1},\tau_j) - x(t_i,\tau_j)$. For $\tau < 3$, $\Delta x(t_i,\tau_j) = 1$ and for $\tau > 3$, $\Delta x(t_i,\tau_j)$ varies according to the slope of the curve $x(t,\tau)$.

The resulting density function $n$ presents three zones separated by two discontinuities (see Figure 4b) at time $t = 6$. For $x < 3$, the density $n$ increases as the slope of the characteristic decreases with the increase in $x$. For $x \in [3,4]$, the density is found by adding the density of populations from both characteristics. For this, we discretize the range of $x(t,\tau)$ by using a new variable $X$ with a constant step size $\Delta X = X_{i+1} - X_i = 0.02$. The step size is chosen to be sufficiently small to allow us to sum up the density functions $n$ between any two nodes $X_i$ to $X_{i+1}$. Finally, the density function for the interval $x > 4$ remains constant as the characteristics are increased with the same slope. The density function with a constant grid of variable X appears in Figure 4b.

*Example 2: Non-monotonous case*

In this example, we illustrate the case of a non-monotonous characteristic presenting crossing from a given value of time. We consider the family of characteristics in the form (see Figure 5a):

$$x(t,\tau) = (t-\tau)a\left(\frac{\exp(-b(t-\tau))}{(\tau+1)}\right),$$

where $a$ and $b$ are constants (here $a = 50$, $b = 0.4$) and $B(t) = 1$, the same as in the previous example. The actual and starting time step size are chosen as $\Delta t = \Delta \tau = 0.04$ while the step size $\Delta x$ varies according to the slope of the curve. The initial and boundary conditions are the same as in the previous example. The density function $n(x,t)$ can also be followed according to the previous example. The constant step size for the discretization of the range of $x(t,\tau)$ has been chosen as $\Delta X = 0.1$.

At short times, before crossing occurs, the density computed by the transport method is a smooth curve, as shown in Figure 5b. The crossing of characteristics begins around time $t = 9$ (Figure 5a), leading to the appearance of a discontinuity in the density curve $n(x, t)$, for the same reasons as exposed in Example 1 (Figure 5c). After a while, the characteristics are grouped into a narrow region of the $(x, t)$ plane (Figure 5a). This behaviour leads to the high values of $n(x, t)$ for the internal variable $x$ as observed in Figure 5c.

APPLICATION TO T-CELLS DYNAMICS

T-cells activation process is a complex dynamic process in which T-cells start producing various proteins, which react with each other to defend themselves against a viral infection.[15] Each T-cell binds with the antigen presenting cell (APC) and recognizes the virus. From the recognition time, i.e. the starting time $\tau$, the T-cell is said to be partially activated. Afterwards, the dynamical changes on the surface of T-cell take place through the down-regulation and the up-regulation of membrane proteins that are measured in their concentration $c$. The process of continuous dynamic change after the starting time $\tau$ is termed as 'activation process' (or process of activation) of a T-cell which lasts until the T-cell is fully activated. Moreover, the actual and starting time step size are chosen to be the same throughout the simulations, $\Delta t = \Delta \tau = 12$.

There are approximately $a^* = 2.16 \times 10^{12}$ T-cells/m$^3$ present in our blood plasma (Table 1). When infection is presented by antigen presenting cells (APCs), all the T-cells do not recognize the infection at once; instead a group of T-cells interact with APCs at each small time step. We assume that at each starting time $\tau$ a group of T-cells interacting with APCs to start their activation process has the same characteristics throughout the simulation time. Therefore, we consider a single T-cell from each group which represents the whole group and follow the variation in its characteristics with time. Each single T-cell in the activation process is characterized by the variation of its proteins concentrations.

Multi-scale mathematical modelling is required to investigate this process. At the first scale, two processes which are termed as the activation and reaction terms (assimilated as nucleation and growth in chemical processes) are studied as independent processes. At the second scale, the dynamical variation in the population density of activated T-cells is investigated in the presence of activation and reaction terms. We discuss the T-cell activation dynamics below for the two scales mentioned above.

**Activation Rate of T-Cells**

The activation rate of T-cells is the rate at which the non-activated T-cells start their activation process. It is denoted by $B(t)$, the number of cells activated per unit of time and per unit of volume ($s^{-1} \cdot m^{-3}$) at $t = \tau$, and can be represented as follows:

$$B(t) = -\frac{dN_{NAC}}{dt} = \frac{dN_{AC}}{dt} = k N_{APC} N_{NAC} = k^* N_{NAC}. \tag{6}$$

Here $k$ is the kinetic constant that indicates the probability of connection between non-activated T-cells and antigen presenting cells (APCs). $N_{APC}$ is the total concentration of APCs loaded with the specific virus and it is assumed constant. The negative sign shows that the population of non-activated T-cells is decreasing. By solving the above differential equation, we obtain the following expression for $B(t)$:

$$B(t) = a e^{-bt}, \quad \text{where } a, b \in \mathbb{R} \tag{7}$$

**Reaction Rates of T-Cells**

At the scale of a single T-cell, three proteins are considered that vary in concentration depending on each other. We consider the rates of change in the protein concentrations to be the reaction rates of T-cells. Therefore, the dynamical variations in the proteins concentrations make a system of ordinary differential equations that can be represented, for a fixed $\tau$, as follows:

$$\frac{dc(t,\tau)}{dt} = \begin{bmatrix} \dot{c} \\ \vdots \\ \dot{c} \end{bmatrix} g(c,t),$$

$$\text{where } \dot{c}_i = \frac{dc_i(t,\tau)}{dt} \text{ and } c(\tau,\tau) = \begin{bmatrix} c_{10} \\ 0 \\ 0 \end{bmatrix},$$

Let us define the model describing the activation process of T-cells. The process, as shown in Figure 6, begins as the CD3 protein gets internalized with a rate that is considered linear with kinetic parameter $k_1$ (s$^{-1}$). We assume that, at every time $t = \tau$, T-cells begin their process of activation with same initial concentration of CD3 protein. The change in the concentration of CD3 protein ($c_1$) is independent of all other proteins and therefore it is represented by the following equation:

$$\dot{c_1} = k_1 c_1(t,\tau), \quad where \quad t > \tau \tag{8}$$

where $c_1(t,\tau) = c_{10}, (t < \tau)$ is the initial concentration of CD3 protein present on the cell surface. The concentration of the internalized CD3 protein (i.e. CD3i) is denoted by $c_2(t,\tau)$ which subsequently increases inside the T-cell. However, the CD3i protein concentration decreases with the kinetic parameter $k_2$ (s$^{-1}$) due to its half-life. In contrast to CD3 protein that is present on the surface of T-cell, CD3i protein lies inside the T-cell, and therefore it is considered a volumetric protein. Therefore we have the following equation:

$$\dot{c_2} = k_1 \frac{S_c}{V_c} c_1(t,\tau) - k_2 c_2(t,\tau), \quad where \quad c_2(t < \tau, \tau) = 0. \tag{9}$$

The parameters $S_c$ and $V_c$ denote the surface and volume of T-cells. The ratio appears because $c_1(t,\tau)$ is a surface concentration while $c_2(t,\tau)$ is a volumetric concentration. At each time $t = \tau$ the T-cells, having CD3i protein concentration, start producing CD25 protein concentration ($c_3(t,\tau)$) on their surfaces with the kinetic parameter $k_3$ (s$^{-1}$). At the same time, the production of Interleukin-II (IL-2) begins and goes into the bloodstream where it is shared by all the T-cells as a messenger. IL-2 protein can also be introduced directly into the bloodstream (by medicine or by injection). By doing so, the quantity of IL-2 protein can be maintained. Moreover, the IL-2 protein is independent of starting time $\tau$ since it is not attached to the T-cells which produce it. In this paper, the IL-2 protein will be assumed to be controlled by an external source and its production rate is assumed as $k_4$ (mol·m$^{-3}$·s$^{-1}$). However, CD25 protein acts like a receptor for IL-2 protein and they combine to make CD25-IL2 complex. For this reason, the concentration of CD25 protein starts decreasing depending on the rate of interaction between CD25 and IL-2 proteins. The interactions happen at the rate defined by the kinetic constant $k_5$ (m$^3$·mol$^{-1}$·s$^{-1}$). Therefore, the rate of change in the CD25 protein concentration takes the following form:

$$\dot{c_3} = \frac{V_c}{S_c} c_2(t,\tau) - k_5 c_3(t,\tau) c_4(t), \quad \text{where } c_3(t < \tau, \tau) = 0, \tag{10}$$

In this article, the IL-2 protein concentration $c_4(t)$ is increased linearly by injection:

$$c_4(t) = k_4 t \quad \Rightarrow \dot{c_4} \tag{11}$$

A set of values is considered for the kinetic parameters $k_i$ ($i$ = 1...5) describing the single T-cell dynamics. The numerical solutions for the system of ODEs (see Equations (8–10)) are solved using the RK-4 method as shown in Figure 7.

The CD3 protein follows a monotonic decrease phenomenon, Figure 7a. The protein concentration curves of two other proteins (CD3i, CD25) show the non-monotonicity as well as crossing phenomena, Figures 7b–c. Moreover, each concentration curve for CD25 protein defines its own path due to the dependence on IL-2 protein.

**3-D Analysis of CD25 Protein Concentration**

To understand the crossing of curves in this problem, we discuss the CD25 protein concentration as shown in Figure 7, by visualizing it in 3-D as shown in Figure 8. The surface plot is cut by three planes which lie at the simulation times $t$ = 3000, 9000, and 15 000 s in Figure 8.

The top-right 2-D plot in Figure 8 contains three curves. Each curve represents the cutting points by the plane which cut the surface at the above three simulation times. The first curve at $t$ = 3000 s shows a monotonic decrease and thus shows that there is no crossing points. The second curve at $t$ = 9000 s shows a small increase in the start that is followed by a decrease afterwards. This means that, for some values of activation time $\tau$, the concentration is repeating itself. These repeated points become the crossing points when the surface plot in Figure 8 is projected on a 2-D plot as shown in Figure 7c. This creates the problem of crossing characteristic curves.

Let us consider that the population is flowing on each characteristic curve in the surface plot in Figure 7c. The population at the crossing point is the sum of two populations travelling on the characteristic curves. Therefore, the population of T-cells gives a drastic change that is observed as discontinuity. However, in general, the reason behind the discontinuity is explained in Example 3. As this is a continuous process, this discontinuity continues for all succeeding curves. This concept is further analyzed in the next section.

**Population Dynamics of Activated T-Cells**

The population density function is found by dividing the characteristic curves into two portions: one without crossing and the second with crossing of curves. To solve the problem, two classical methods (method of characteristics (MOC)[29] and Lax-Wendroff finite difference scheme with flux limiter (LWF)[19,40]) are used to investigate the population dynamics for the surface protein concentrations (CD3 and CD25). Regardless of numerical techniques, the objective is to show that the population density cannot be found on the crossing characteristics by using PBMs which follow the linear hyperbolic conservation equation. Therefore classical numerical techniques cannot be applied to find the solution at the crossing points. The newly developed transport method (TM) works in this case and provides a smooth solution to the crossing points in contrast to the classical techniques.

The time steps are chosen to be the same as defined in the start of the application section while the concentration step size varies according to the slope of the characteristic curves. In cases without crossing monotonic curves, as in Figure 9, we discretize the $c_1$ as variable $C_1$ and set the step sizes $\Delta C_1(t_i) = c_1(t_{i+1}, 0) - c_1(t_i, 0)$ since all the curves are translated from the first curve (i.e. when $\tau = 0$). The initial condition is set $n(c,0) = 0$ since it is considered that no cell is activated at time $t = 0$. The boundary condition is derived from the fact that T-cells start their activation process when they recognize the virus. Therefore, we take the ratio between the rate at which cells start their activation process (activation rate $B(t)$) and the rate at which the initial concentration of CD3 protein decreases (reaction rate $G(c_1 = c_{10})$). We write it as follows:

$$n(c_{10}, t) = \frac{B(t)}{G(c_{10})}$$

In this case without crossing, TM and LWF have overlapping behaviour with MOC to find the population density of activated T-cells as shown in Figure 9 (CD3 protein). In Figure 10, the classical numerical method (e.g. LWF) is applied with difficulty in the concentration grid-points since these are not the same for each concentration curve, as seen in Figure 7c. Additionally, the Courant-Friedrichs-Lewy (CFL) condition cannot be satisfied. Therefore, it is mandatory to interpolate the grid-points and make a fixed concentration axis for all the curves before applying the classical numerical technique LWF. The step size for the fixed concentration grid-points, say $C_3$, can be followed in the same way as defined in above examples. We take $\Delta C_3(t_i,\tau_j) = c(t_{i+1},\tau_j) - c(t_i,\tau_j)$. This generates errors, observed in the form of diffusion in Figure 10. In contrast, TM is independent of the CFL condition and therefore is not obliged to use the interpolation before finding the population density function.

The classical methods work quite well for all portions of curves in which the characteristics do not cross each other. A relative error (RE) analysis was studied to validate numerical schemes, using the following formula:

$$RE = \left| \frac{M_1 - M_2}{M_1} \right|$$

In Figure 11a, the relative error of Figure 9 is investigated by choosing $M_2$ once as LWF and then as TM while $M_1$ is chosen as the analytical solution that is found by using MOC at the final time $t = 2.4 \times 10^4$ s. Both numerical methods agree closely with MOC and present less error. TM error remained consistent throughout the simulation because it does not depend upon the finite differences/volumes of the neighbouring concentrations, while LWF decreases as the concentration increases. It is important to mention here that the concentration grid is variable (i.e. first curve $c_1(t, 0)$), having more grid points at the start and gradually decreasing. Since we do not have an analytical solution in Figure 10, we plotted relative error between LWF as $M_1$ and TM as $M_2$ in Figure 11b. At the portion without crossing characteristics, LWF and TM work quite well even though LWF uses a fixed grid for the concentration $c_3$ axis as defined above. However, at the portion of crossing curves, as shown in Figure 7c, two populations acquire the same concentration which cannot be followed by classical methods, e.g. LWF, and therefore generates the error against TM (Figure 11b). In the classical methods, the population density at any time $t$ is found by following the flux of the population at the previous time step while the characteristics were not crossing each other. However, given the crossing of characteristics problem, we need two fluxes at this point, since the slope of each characteristic is different. Therefore, the advection equation cannot be written. For the same slopes of crossing characteristic curves, the model can be represented by couple of population balance equations.[35] Since the activation process is continuous, the crossing will be observed continuously as well. Thus, the population balance equation cannot be written correctly for this portion as shown in Figures 10 and 11b. One can observe it easily for $t \geq 9000$ s, where the LWF does not hold for the region of crossing curves. However, TM finds the population density efficiently.

CONCLUSION

This article studied non-conventional growth laws when the characteristics cross each other. A new method is developed to find the population density function for multiscale mathematical modelling. The transport method showed close agreement with the classical methods to find the density function when the concentration curves did not cross each other. However, the crossing behaviour in the concentration (internal variable) curves did not allow us to write the classical conservation equation.

The developed transport method can find the population density function only when the reaction and activation processes are involved. However, the method has the capacity to be modified to study the population density function when other physical (particulate) processes take place. For continuous distribution $D(x)$ of the activation process term when the boundary condition cannot be extracted from the source term, the method would not work in its current formulation since every point on the $(x, t)$ plane would have an initial population density that needs to be added into the population flow. In other words, every point on the $(x, t)$ plane would be a starting point of a new characteristic curve over which the initial population exists while another population flowing over any preceding curve may cross it. When this process would be continuous, there may not be any discontinuity observed in the population density function, contrary to what we observed in Figure 10 (population density shows a discontinuity at the very first point where the curves cross each other). However, the classical way of writing PDEs for such systems would remain a challenge. In addition, the method would be very helpful for studying the multidimensional case when the population density function involves two or more internal variables in the dynamical system.

T-cells immune response is based upon a very complex signalling process which comprises hundreds of proteins and produces intracellular as well as extracellular signalling. The above model is simple compared to the complexity of the activation process; however, it is not necessary to follow all the proteins at once since it is possible to study groups of proteins which depend upon each other. This can help us understand the specific behaviours of T-cells immune response. Moreover, the activation process of T-cells depends only upon the reaction and activation rates, which are already considered in the transport method, while most of the protein concentrations like CD25 are produced by T-cells from almost zero which allows us to make boundary conditions for their population densities. However, CD3 protein concentration varies from cell to cell which makes for a continuous initial distribution $D(x)$ of CD3 T-cells instead of its current initial distribution of $\delta$-function. Last but not least is the insertion of the delay parameter for the production of proteins since the activation process of T-cells does not begin to produce all the proteins from the very beginning. For example, CD3 protein triggers the activation process and continues to down-regulate for a few hours and strengthen the cell to produce more protein such as CD25 and IL-2.

**NOMENCLATURE**

| | |
|---|---|
| $a$ | number of interactions between $N_{APC}$ and $N_{NAC}$ (m$^{-3}$·s$^{-1}$) |
| $a^*$ | initial density of $N_{NAC}$ (m$^{-3}$) |
| $b$ | volumetric flow rate of $N_{APC}$ (s$^{-1}$) |
| $B(t)$ | activation rate of T-cells (m$^{-3}$·s$^{-1}$) |
| $c_1(t, \tau)$ | surface concentration of CD3 (mol·m$^{-2}$) |
| $c_2(t, \tau)$ | volume concentration of CD3 (mol·m$^{-3}$) |
| $c_1'(t, \tau)$ | surface concentration of CD3* (mol·m$^{-3}$) |
| $D(c)$ | initial distribution of non-activated T-cells (mol$^{-1}$·m$^{2}$) |
| $G(c,t,\tau)$ | reaction rate of protein concentration (mol·m$^{-2}$·s$^{-1}$) |
| $k^*$ | volumetric contact rate (m$^{3}$·s$^{-1}$) |
| $k_1$ | rate of CD3 internalization (s$^{-1}$) |
| $k_2$ | rate of CD3i diminution (s$^{-1}$) |
| $k_3$ | production rate of CD25 (s$^{-1}$) |
| $k_4$ | production rate of IL2 (mol·m$^{-3}$·s$^{-1}$) |
| $k_5$ | interaction rate of CD25 with IL-2 (mol$^{-1}$·m$^{3}$·s$^{-1}$) |
| $n(c,t)$ | density of T-cells relative to surface protein (mol$^{-1}$·m$^{-1}$) |
| $N_{AC}(t)$ | density of activated T-cells (m$^{-3}$) |

$N_{APC}$     density of antigen presenting cells $(m^{-3})$

$N_{NAC}(t)$   density of non-activated T-cells $(m^{-3})$

$N(c,t)$     initial distribution of activated T-cells $(mol^{-1} \cdot m^{-1} \cdot s^{-1})$

t and τ     actual time and activation time (s)

$S_c$        surface area of a T-cell ($m^2$)

$V_c$        volume of a T-cell ($m^3$)

CAPTIONS OF THE FIGURES

Figure 1: Family of characteristic curves $x(t,\tau)$. $B(t)$ gives a distribution of the initial population (dashed line). $n(x,t_i)$ is the density function at time $t = t_i$ and variable $x$.

Figure 2: Crossing of two characteristic curves $x_A$ and $x_B$ in $(x,t)$ plane showing non-monotonic behaviour and followed by two different initial populations: $P_A = B(\tau_A)\Delta\tau$ and $P_B = B(\tau_B)\Delta\tau$. Population densities for each curve $x_A$ and $x_B$ at the point of crossing are $n(x_{A,i}, t_{j+1})$ and $n(x_{B,i}, t_{j+1})$ respectively, where $x_{A,i} = x_{B,i}$.

Figure 3: Variation in the initial population with starting time between $\tau_i$ and $\tau_i + \Delta\tau$ and its relation with population density at time $t_j$ and internal variable between $x_i$ to $x_i + \Delta x_i$.

Figure 4: (a) Characteristic curves with slopes $g_\tau(x(t,\tau), \tau) = 1$ for $\tau < 3$ and $g_\tau(x(t,\tau), \tau) = 1 + \tau - 3$ for $\tau \geq 3$. Horizontal axis represents actual time $t$ and delay time $\tau$ whereas vertical axis represents variable $x$. Horizontal lines at $x = 3$ and $x = 4$ capture the crossing portion of vertical lines at $t = 6$. (b) Discontinuity in the density function at time $t = 6$ is plotted when the curves are crossing between $x = 3$ and $x = 4$.

Figure 5: (a) Characteristic curves in $(x, t)$ plane where each curve is following an independent path. (b) Solution by transport method in $(n, x)$ plane at time $t = 3$ (dashed line), 6 (dashed dotted line), and 9 (solid line). (c) Solution by transport method in $(n, x)$ plane at time $t = 12$ (dotted), 15 (dashed), 18 (dashed dotted), and 21 (solid).

Figure 6: Activation process of T-cells described in three steps. Step 1: Internalization of CD3 protein. Step 2: Production of CD25 protein due to internalization of CD3 protein. Step 3: Binding of IL-2 protein to its receptor CD25 protein. The variable in the figure represents the protein concentration. $c_1$ is the surface protein CD3 concentration, $c_2$ is the CD3i protein concentration (CD3 internalized into T-cell), $c_3$ is the CD25 surface protein, and $c_4$ is the IL-2 protein present in blood plasma outside the T-cell and shared by all T-cells.

Figure 7: Protein concentrations versus time. Graphical representation of the solution of the above system of ODEs with the parameteric values $k_1$, $k_2$, $k_3$, $k_4$, and $k_5$ are mentioned in Table 1.

Figure 8: 3-D visualization of CD25 protein concentration using ($\tau$,t,c) surface plot. The surface plot is cut by three planes at time t = 3000, 9000, 15 000 s. Cross-sections of the planes at their intersection with surfaces are shown in 2-D at the top right of this figure with axis ($\tau$,c).

Figure 9: Population density of CD3 protein at 4 different times in seconds: (a) t = 3000, (b) t = 9000, (c) t = 15 000, and (d) t = 24 000. Horizontal axis of each figure is the concentration 'c' while the vertical axis is the population density n(c,t). The solution was found by three methods: TM = straight lines, method of characteristics = crosses, and Lax-Wendroff numerical simulations = squares. Parametric values are mentioned in Table 1.

Figure 10: Population density for CD25 protein at 4 different times in seconds: (a) t = 3000, (b) t = 9000, (c) t = 15 000, and (d) t = 24 000. Horizontal axis of each figure is the concentration 'c' while the vertical axis is the population density n(c,t). The solution found by TM = straight lines; by Lax-Wendroff numerical scheme = squares. Kinetic parameters are mentioned in Table 1.

Figure 11: Relative error analysis of Transport Method and Lax-Wendroff scheme. (a) TM (dashed line) and LWF (solid line) are validated against MOC for the population density of T-cells in the process of activation with respect to CD3 protein concentration as shown in Figure 9. Error is estimated at the final time $t$ = 24 000 s. (b) In the absence of analytical solution, relative error was studied between TM and LWF at four different values of t on which the population density of activated T-cells with respect to CD25 protein concentration is plotted in Figure 10.

CAPTIONS OF THE FIGURES

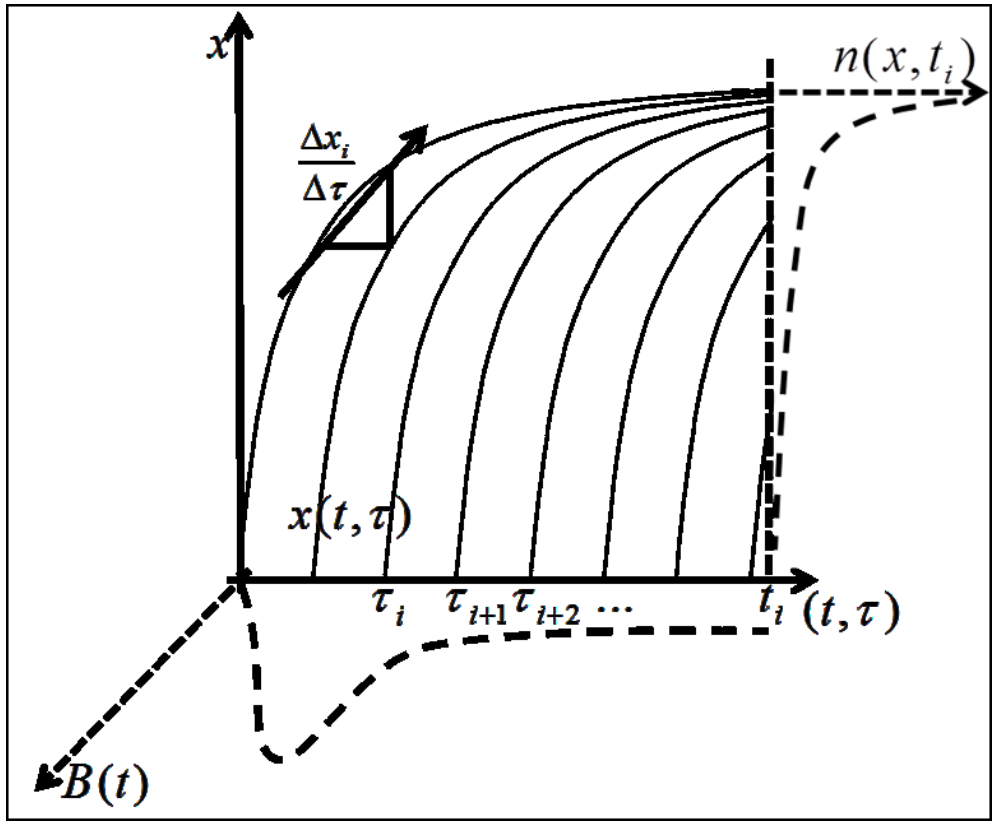

Figure 1: Family of characteristic curves $x(t,\tau)$. $B(t)$ gives a distribution of the initial population as shown by dashed line. $n(x,t_i)$ is the density function at time $t = t_i$ and variable $x$.

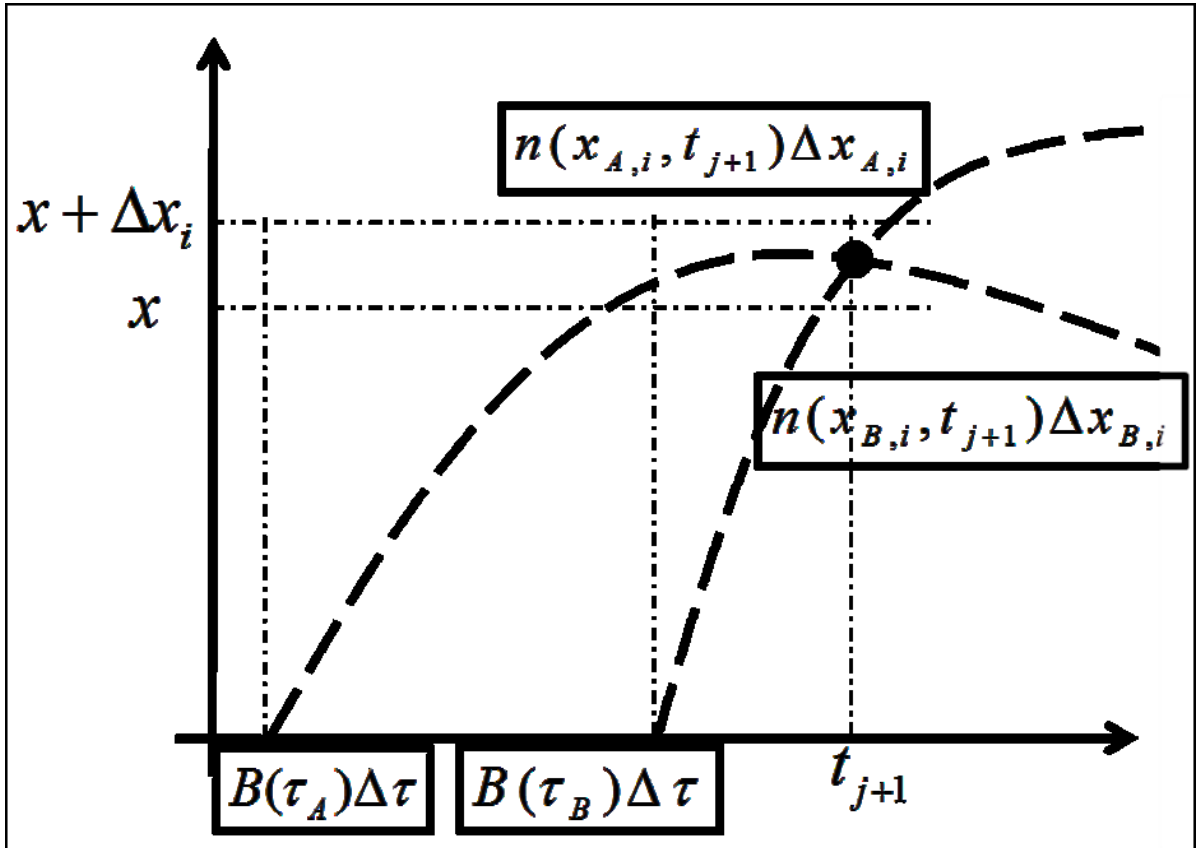

Figure 2: Crossing of two characteristic curves $x_A$ and $x_B$ in $(x,t)$ plane showing non-monotonic behavior and followed by two different initial populations $P_A = B(\tau_A)\Delta\tau$ and $P_B = B(\tau_B)\Delta\tau$. The population densities for each curve $x_A$ and $x_B$ at the point of crossing are $n(x_{A,i},t_{j+1})$ and $n(x_{B,i},t_{j+1})$ respectively, where $x_{A,i} = x_{B,i}$.

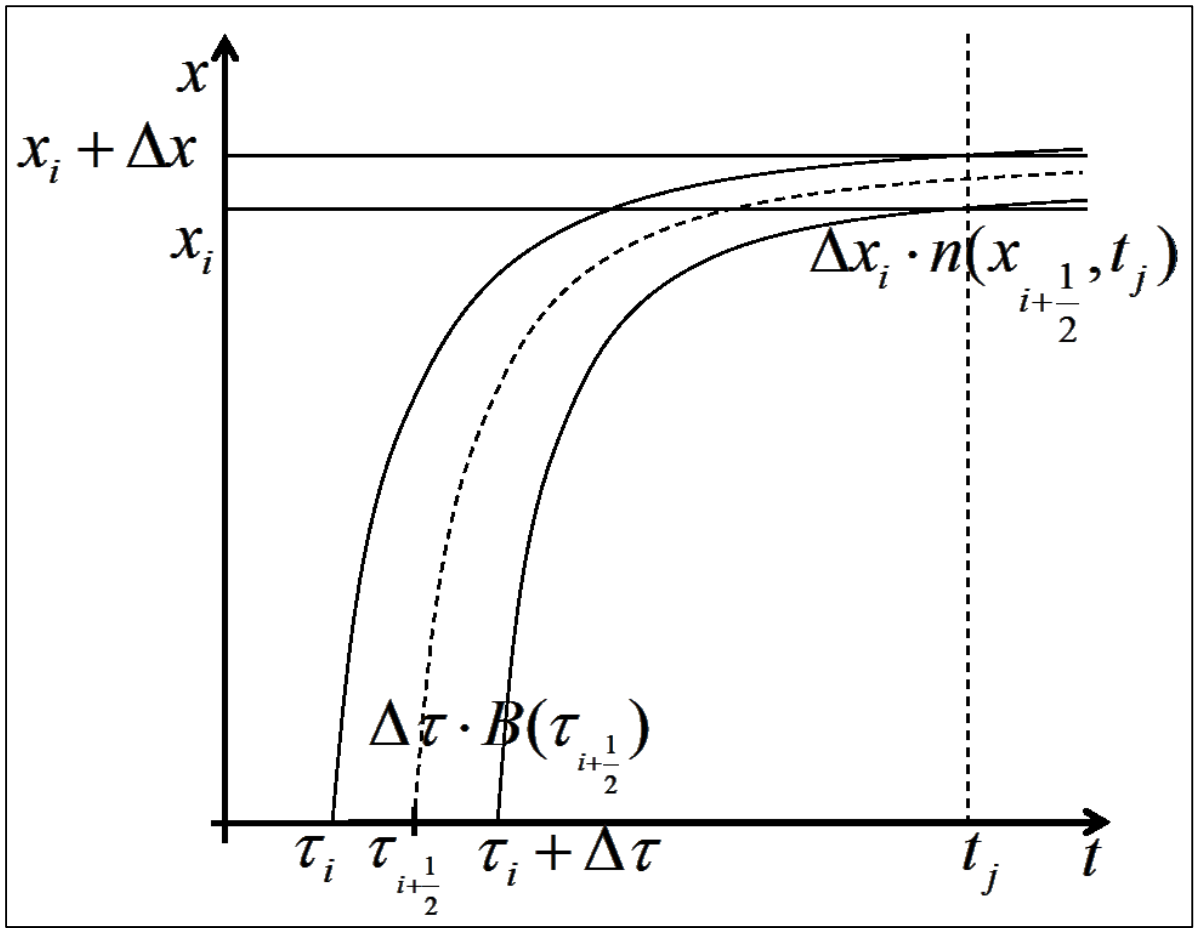

Figure 3: Variation in the initial population with starting time between $\tau_i$ and $\tau_i + \Delta \tau$ and its relation with population density between the time $t$ to $t + \Delta t$ and $x_i$ to $x_i + \Delta x_i$.

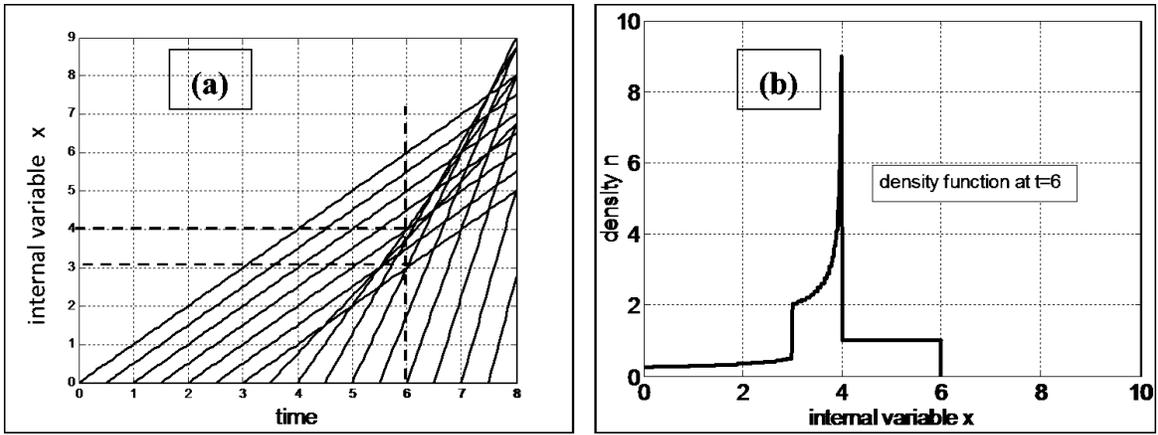

Figure 4: $g(t,\tau)=1$ for $\tau<3$; $g(t,\tau)=1+\tau-3$ for $\tau\geq 3$; **(a)** characteristic curves; **(b)** Discontinuity in the density function at time $t=6$ when the curves are crossing for values $3<x<4$.

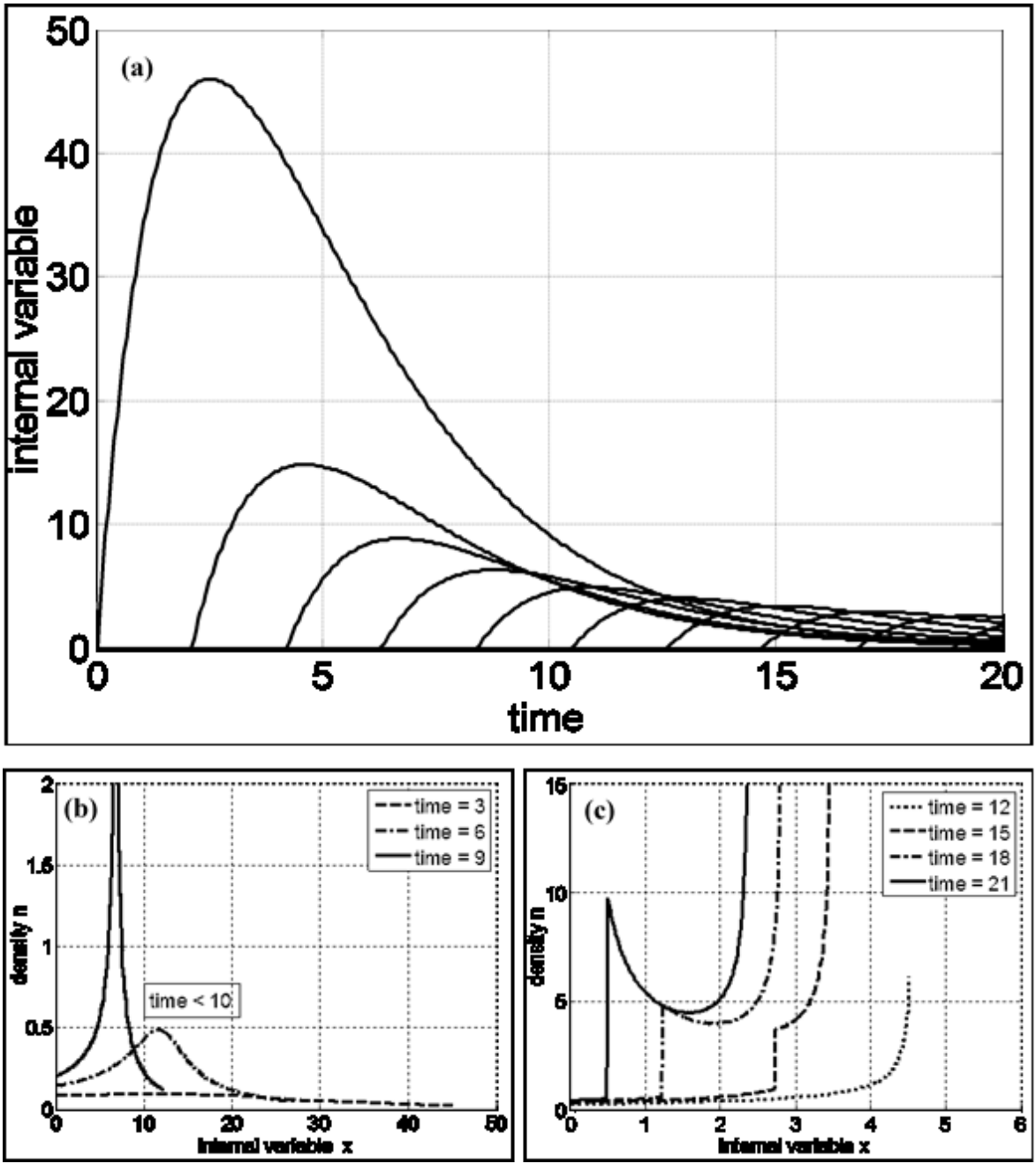

Figure 5: (a) Characteristic curves in (*x, t*) plane. (b) Solution by transport method at time *t* = 3, 6, 9. (c) Solution by transport method at time *t* = 12, 15, 18, 21.

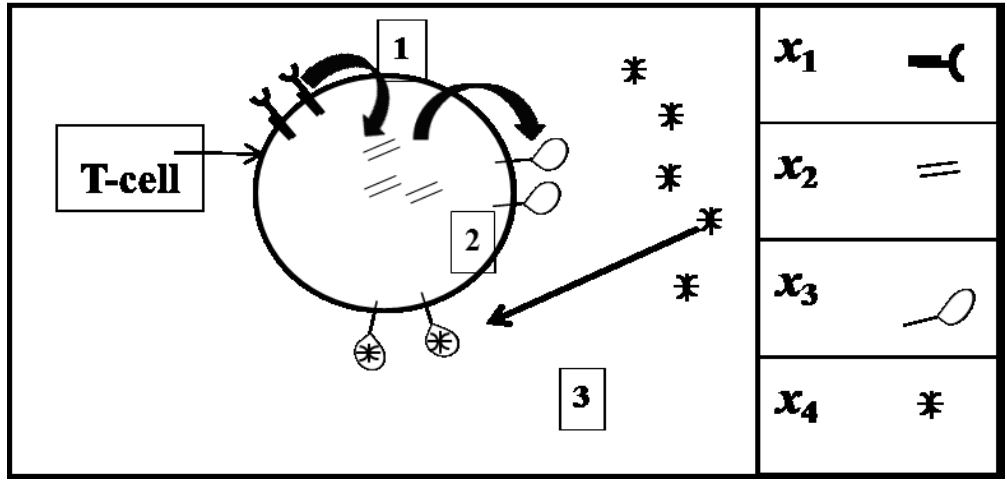

Figure 6: Activation process of T-cells described in three steps. Step 1: Internalization of CD3 Protein. Step 2: Production of CD25 protein due to internalization of CD3 protein. Step 3: Binding of IL-2 protein to its receptor CD25 protein. The variable in the figure represents the protein concentration. $c_1$ is the surface protein CD3 concentration, $c_2$ is the CD3i protein concentration (CD3 internalized into T-cell), $c_3$ is the CD25 surface protein and $c_4$ is the IL-2 protein present in blood plasma outside T-cell and shared by all T-cells.

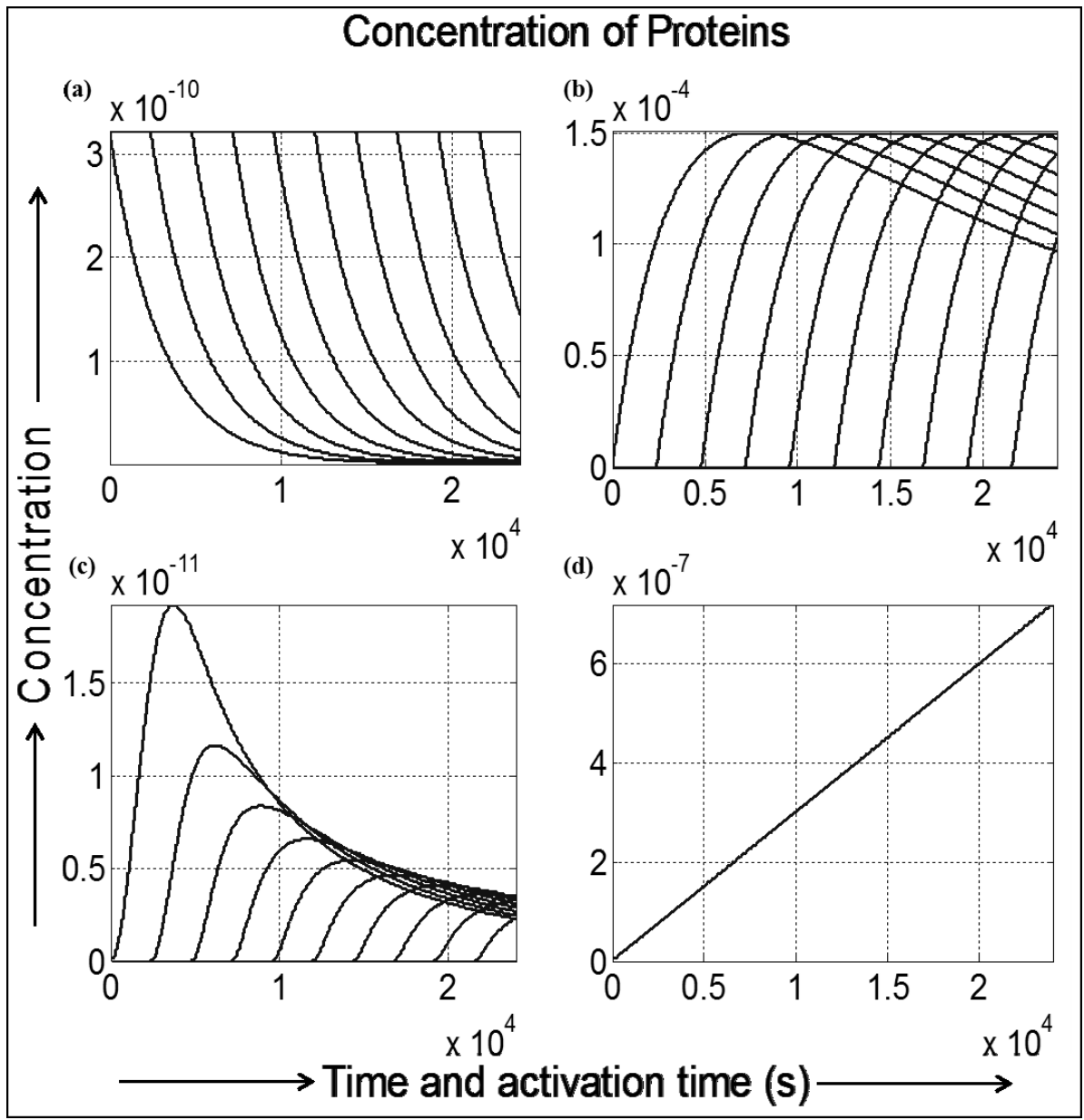

Figure 7: Protein concentrations versus time. Graphical representation of the solution of above system of ODEs with the parameter values:

$k_1 = 1/3 \cdot 10^{-3} s^{-1}$, $k_2 = 1/3 \cdot 10^{-4} s^{-1}$, $k_3 = 10^{-4} s^{-1}$, $k_4 = 3 \cdot 10^{-11} mole \cdot m^{-3} s^{-1}$, $k_5 = 10^4 m^3 mole^{-1} s^{-1}$.

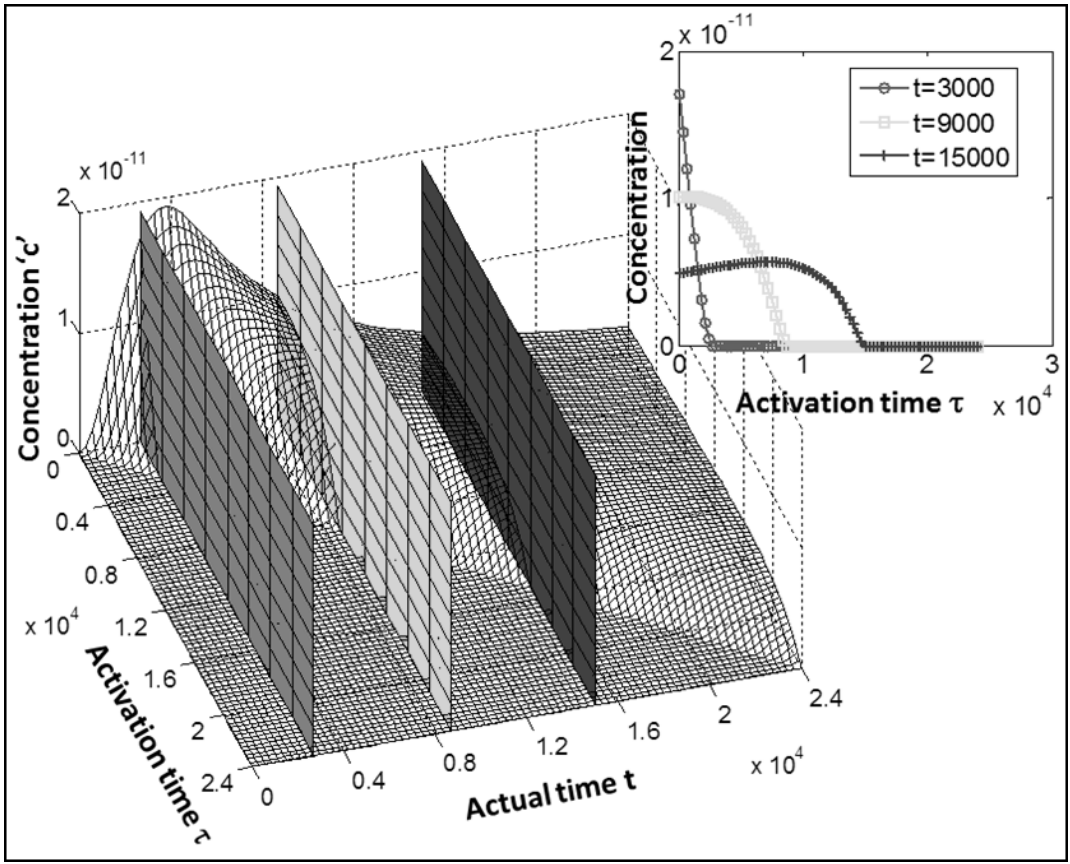

Figure 8: 3D visualization of CD25 protein concentration cut by planes. The cutting curves are shown in the 2D plots attached at the top right of each figure.

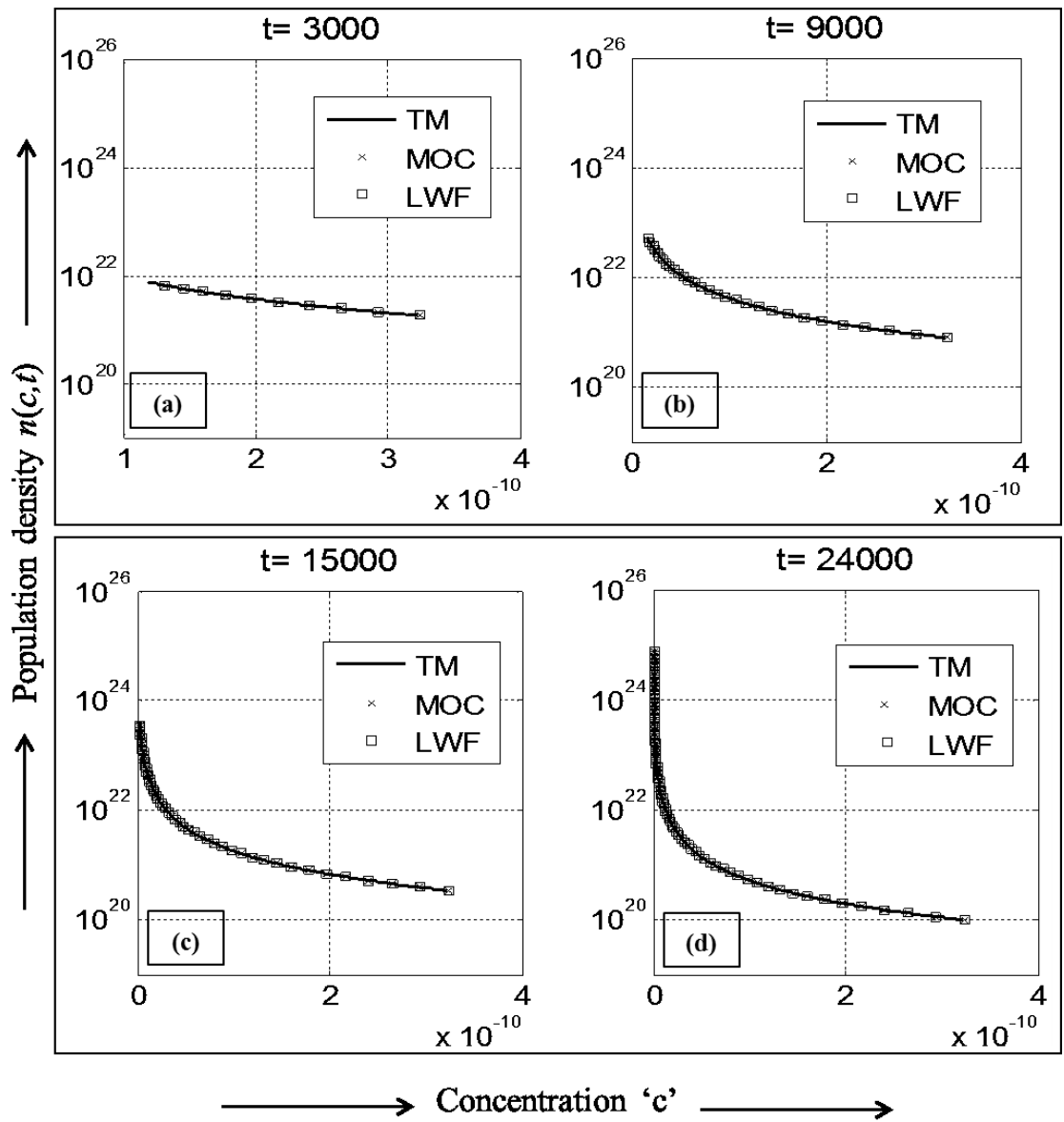

Figure 9: Population density for CD3 protein at 4 different times in seconds. The kinetic parameter for CD3 protein is $k_1 = 1/3 \cdot 10^{-3} s^{-1}$.

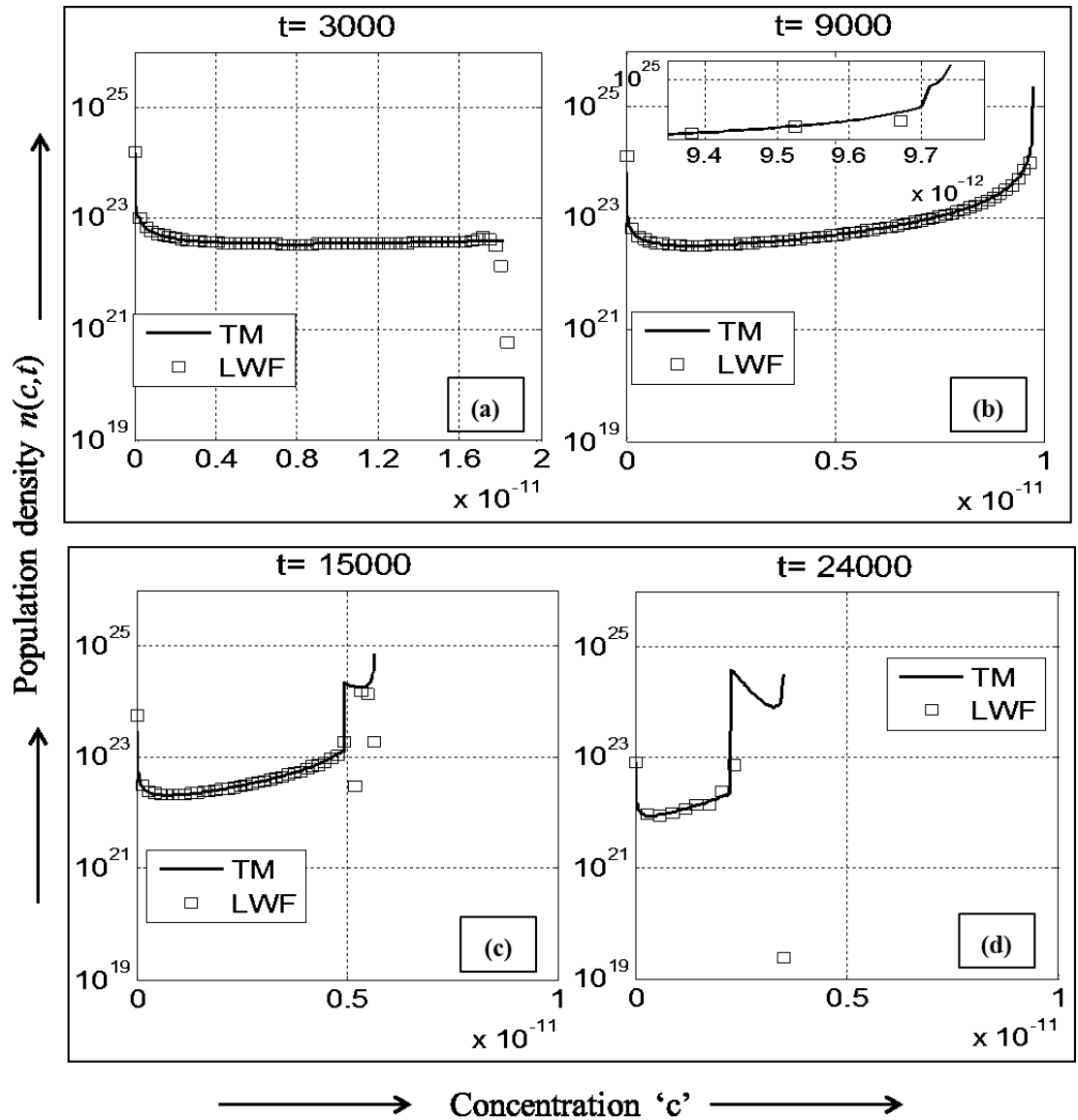

Figure 10: Population density for CD25 protein at 4 different times in seconds. Kinetic parameters are:
$k_1 = 1/3 \cdot 10^{-3} s^{-1}$, $k_2 = 1/3 \cdot 10^{-4} s^{-1}$, $k_3 = 10^{-4} s^{-1}$, $k_4 = 3 \cdot 10^{-11} mole \cdot m^{-3} s^{-1}$, $k_5 = 10^4 m^3 mole^{-1} s^{-1}$

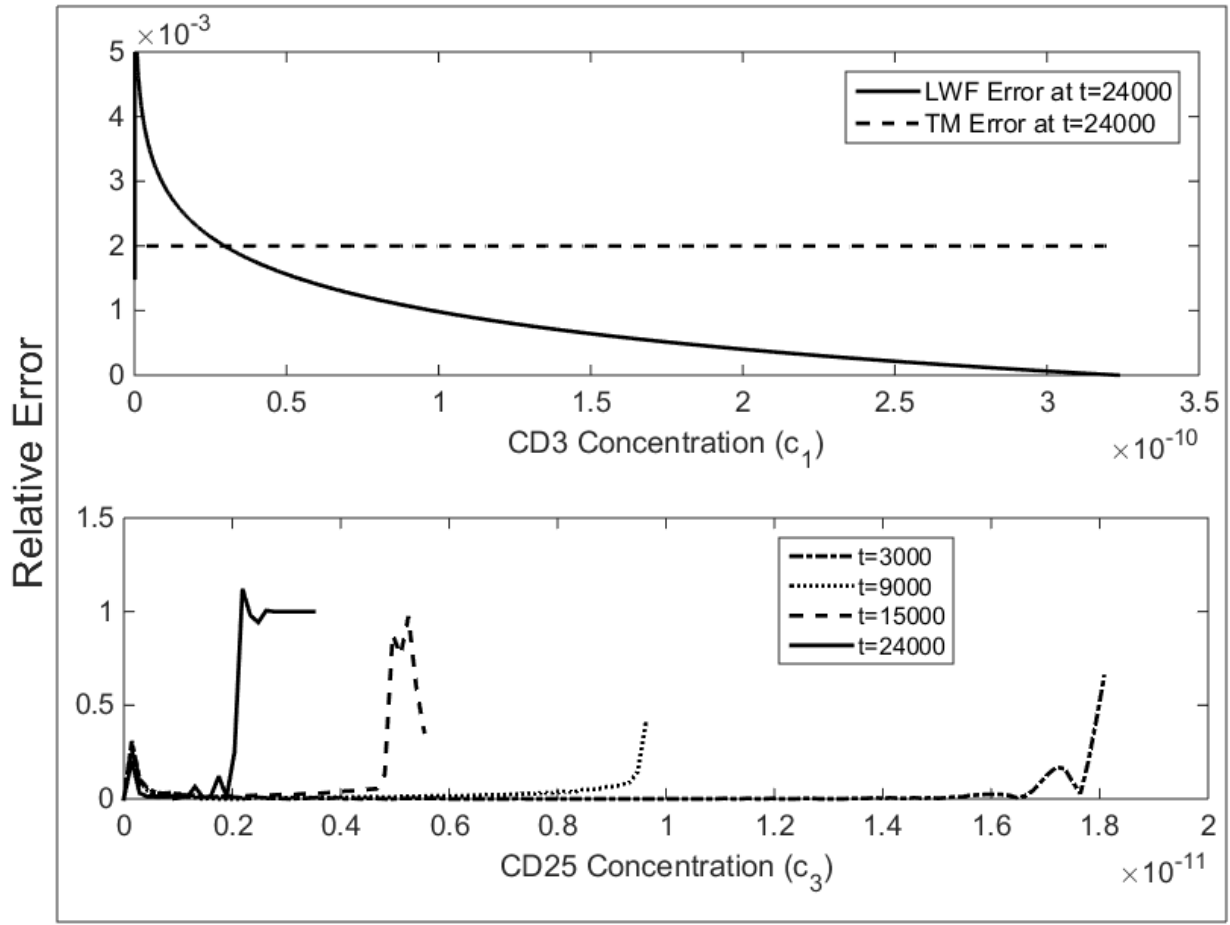

Figure 11: Relative error analysis of Transport Method and Lax Wendroff scheme. (a) TM (dashed line) and LWF (solid line) are validated against MOC for the population density of T-cells in the process of activation with respect to CD3 protein concentration as shown in Figure 9. The error is estimated at the final time $t = 24000\ s$. (b) In the absence of analytical solution, relative error has been studied between TM and LWF at four different values of t on which the population density of activated T-cells with respect to CD25 protein concentration is plotted in

Figure 10.